\documentclass[twocolumn,showpacs,preprintnumbers,amsmath,amssymb]{revtex4}

\usepackage{graphicx}
\usepackage{dcolumn}
\usepackage{bm}

\begin{document}


\title{Commensurate-Incommensurate Magnetic Phase Transition in Magnetoelectric Single Crystal LiNiPO$_4$}

\author{D. Vaknin,$^1$ J. L. Zarestky,$^1$ J.-P. Rivera,$^2$ and H. Schmid$^2$}
\affiliation{$^1$Ames Laboratory and Department of Physics and Astronomy Iowa State University, Ames, Iowa 50011\\
$^2$Department of Inorganic, Analytical and Applied Chemistry, University of Geneva, Sciences II, 30 quai E. Ansermet, CH-1211-Geneva 4, Switzerland
}%


\date{\today}

\begin{abstract}
Neutron scattering studies of single-crystal LiNiPO$_4$ reveal a spontaneous first-order commensurate-incommensurate magnetic phase transition.   Short- and long-range incommensurate phases are intermediate between the high temperature paramagnetic and the low temperature antiferromagnetic phases.     The modulated structure has a predominant antiferromagnetic component, giving rise to satellite peaks in the vicinity of the fundamental antiferromagnetic Bragg reflection, and a ferromagnetic component giving rise to peaks at small momentum-transfers around the origin at $(0,\pm Q,0)$.  The wavelength of the modulated magnetic structure varies continuously with temperature. It is argued that the incommensurate short- and long-range phases are due to spin-dimensionality crossover from a continuous to the discrete Ising state.    These observations explain the anomalous first-order transition seen in the magnetoelectric effect of this system.
\end{abstract}

\maketitle
Common magnetic systems with simple colinear long-range-order ground state can {\it melt} into the paramagnetic (disordered) state  directly, usually via a second order phase transition, or through a series of intermediate spatially modulated phases before losing all correlations\cite{Ostlund81,Bak82,Kocinski90}.  The indirect melting through modulated phases indicates the presence of competing interactions of next nearest-neighbors, anisotropies in the spin Hamiltonian, and/or topological frustrations\cite{Bak82}.   There has been a continuous interest in the spontaneous and magnetic-field induced commensurate-incommensurate magnetic (C-IC) transition over the years\cite{Chattopadhyay86,Kiryukhin96,Zheludev97}.   For instance, in the semimetallic europium tri-arsenide (EuAs$_3$), the ground state of the system is commensurate and as temperature increases the system undergoes a C-IC transition\cite{Chattopadhyay86}.  In copper metaborate, on the other hand, the ground state is incommensurate and undergoes a continuous phase transition to a non-collinear commensurate antiferromagnetic state\cite{Roessli2001}.   It has also been demonstrated that the C-IC transition can be induced by the application of an external magnetic field\cite{Kiryukhin96,Zheludev97}.   

Here, we report a novel C-IC magnetic phase transition in the weakly coupled antiferromagnetic planes of LiNiPO$_4$ (S=1, Ni$^{2+}$), its characteristics resemble IC structural phase transitions\cite{Janssen82}.
\begin{figure}
\includegraphics{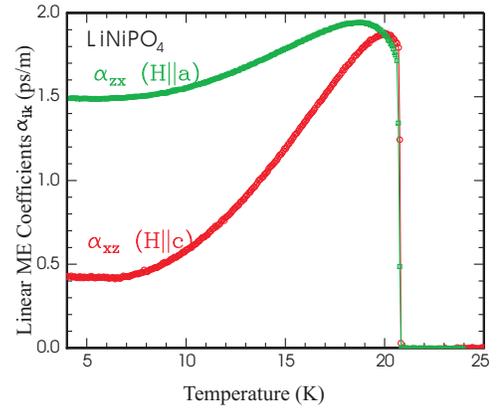}
\caption{\label{ME}  Magnetoelectric coefficients of LiNiPO$_4$ versus temperature measured by the dynamic technique. The ME coefficients, $\alpha_{xz}$, and $\alpha_{zx}$, were measured under 5 kOe  magnetic field along the c-axis and the a-axis, respectively\cite{Rivera94}.}
\end{figure}
LiNiPO$_{4}$ is an antiferromagnetic (AF) insulator\cite{Mays63,Santoro66} which belongs to the olivine family of lithium orthophosphates LiMPO$_{4}$ (M = Mn, Fe, Co, and Ni)\cite{Megaw73}; space group is {\it Pnma}\cite{Abrahams93}.   Neutron scattering studies demonstrated that  LiMPO$_{4}$ (M =Ni, Co) exhibit properties between the two-dimensional (2D) and three-dimensional (3D) with an interlayer coupling that is relatively stronger than the coupling found in the cuprates, for instance\cite{Vaknin99,Vaknin2002}.  These insulators also exhibit a strong linear magnetoelectric (ME) effect, with the observed ME tensor components, $\alpha_{xy}, \alpha_{yx}$, for LiCoPO$_4$ and, $\alpha_{xz}, \alpha_{zx}$, for LiNiPO$_4$, in agreement with the antiferromagnetic point groups mmm' and mm'm, respectively, but with some anomalies\cite{Mercier71a,Mercier71b,Rivera94}.  In particular, the ME effect measurements of LiNiPO$_{4}$ as a function of temperature reveal a first-order AF transition, and an unusual decrease of the ME coefficient at temperatures below a maximum close to $T_N$\cite{Mercier68}.  By contrast, the isostructural LiCoPO$_4$, LiFePO$_4$, and LiMnPO$_4$ exhibit continuous change of the ME coefficients, indicative of second-order transitions\cite{Mercier71a}.    Magnetic susceptibility studies of polycrystalline LiNiPO$_4$ showed a significant deviation from the Curie-Weiss law in a temperature range much higher than $T_N$, and neutron scattering from the same polycrystalline sample gave rise to diffuse scattering at the nominal position of the AF Bragg reflection up to $T \approx 2T_N$\cite{Vaknin99}.  Recent magnetic susceptibility measurements of single crystal LiNiPO$_4$ showed two features, one at T$_N$ = 20.8 K  and one at T$_i$ = 21.8 K associated with an AF transition and an intermediate-phase, speculated to be IC, respectively\cite{Kharchenko2003}. 


An irregular shaped single crystal (0.396 grams in size; lattice constants at RT: $a = 10.0317$ {\AA}, $b = 5.8539$ {\AA},$c = 4.6768$ {\AA}), synthesized by a flux method described elsewhere \cite{Fomin2002}, was used for the neutron scattering studies.     Neutron scattering measurements were carried out on the HB1A triple-axis spectrometer at the High Flux Isotope Reactor (HFIR) at Oak Ridge National Laboratory.  A monochromatic neutron beam of wavelength $\lambda $ = 2.368 \AA\ (14.7 meV, $k_{o}=2\pi /\lambda =2.653$\AA $^{-1}$) was selected by a double monochromator system, using the $(0,0,2)$ Bragg reflection of highly oriented pyrolytic graphite (HOPG) crystals.  The $\lambda /2$ component in the beam was removed (to better than 1.3 parts in 10$^{6}$) by a set of HOPG crystals situated between the two monochromating crystals.  The collimating configuration $40^{\prime },40^{\prime },Sample,34^{\prime },68^{\prime }$ was used throughout the experiments.  HOPG was also used as the analyzer crystal. Temperature measurements and control were achieved by a Conductus LTC-20 temperature controller using Lake Shore silicon-diode temperature sensors (standard curve 10).  ME effect measurements on thin polished plates with evaporated gold layers as electrodes were performed with the dynamic technique and quasistatic technique for calibration, as described elsewhere\cite{Rivera94}. Figure\ \ref{ME} shows a strong ME effect with an abrupt transition, with an anomalous temperature dependence of the ME coefficients, $\alpha_{zx}$ and $\alpha_{xz}$, with a maximum close to T$_{\rm{N}}$, in agreement with previous results\cite{Mercier68,Mercier71a}.   
\begin{figure}
\includegraphics[width=3.2 in]{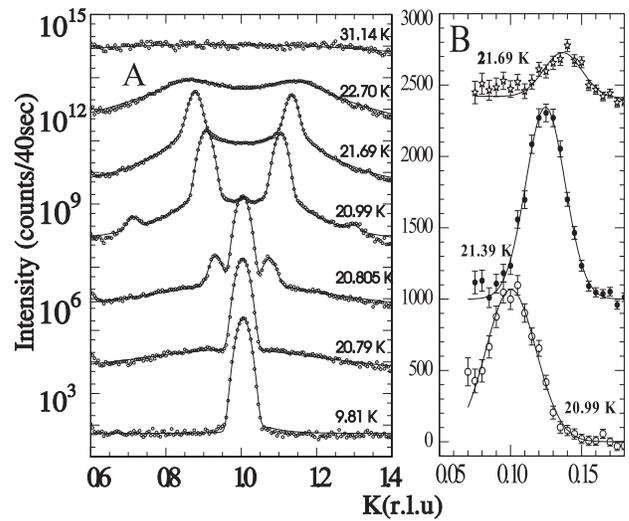}
\caption{\label{Long_scans}  A) Longitudinal scans along the (0,K,0) direction showing a single sharp (resolution limited) peak at the (010) position (r.l.u is reciprocal lattice unit, in this case, b$^*$ = 1.0783 {\AA}$^{-2}$ units; the intensities of scans are shifted by two decades each for clarity).  A single Bragg reflection, due to the  AF ordering, is observed at low temperatures (2 - 19 K).  At and above T$_{\rm{N}}$ long-range IC order predominates and a third order reflection is also observed as shown for T=20.99 K.  Above T $\approx$ 21.7 K, broader peaks associated with the IC  are observed up to $\approx$ 36 K.  B)  Longitudinal scans, close to the origin, along the (0,K,0) direction show a peak compatible in position with the IC peaks observed at (0,1$\pm Q$,0).  This peak, due to a ferromagnetic component of the modulated spin structure, is consistent with Model-I as described in the text (scans shown were obtained after subtraction of similar scans taken at 40K; scans are shifted in intensity for clarity). }
\end{figure}
\begin{figure}[hb]
\includegraphics[width=3.2 in]{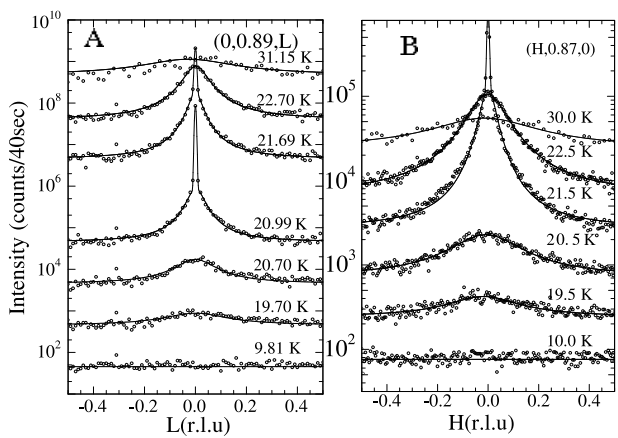}
\caption{\label{Trans_scans}  
A) Transverse scans along L B) and along H  at roughly the position of the IC peak.  In the temperature range 20.8 - to 21.7 K (i.e., from T$_{\rm{N}}\equiv$ T$_{C-IC}$ to T$_{\rm{IC}}$) a broad diffuse peak is superimposed on a resolution limited peak (both Lorentzians). Short range IC phase persists up to T$_{\rm{CO}}\approx$ 36 K.}
\end{figure}

Relatively wide longitudinal scans along the (0,K,0) direction at selected temperatures show the transition from the AF to the paramagnetic state proceeds through an infinite series of modulated structures.  At low temperatures (below T $\approx$ 19 K), a single Bragg reflection at ${\bm\tau}_{_{AF}} \equiv$ (010), due to the colinear AF ordering\cite{Santoro66}, is observed, as shown in Fig.\ \ref{Long_scans} for T = 9.81 K.  At temperatures higher than $\approx 19$ K and lower than $T = 20.80 $ K, this Bragg reflection is superimposed on a very weak diffuse scattering in the form of a broad Lorentzian-shaped peak, also centered at (0,1,0).  This broad peak (width 0.235 {\AA}$^{-1}$), due to short-range inplane coherence-lengths ($\xi \approx$ 27 {\AA}), is likely related to the lamellated domains recently observed by magnetic second-harmonic generation topography on a thin plate of LiNiPO$_4$\cite{Leute2001}, and with the maximum observed in the ME effect(see Fig.\ \ref{ME}).  At T = 20.80 K, two extra satellite reflections at (0,1$\pm$Q,0) appear, signaling a transition from the simple colinear AF phase to the IC magnetic phase.  The transition occurs within 0.005 K of $T_{\rm{N}}$ (the resolution in temperature). As shown, at this temperature commensurate and incommensurate phases coexist, as is typical of first-order phase transitions and consistent with the sudden disappearance of the linear ME effect (former observations show that the linear ME-effect cancels out in magnetically incommensurate structures; e.g. BiFeO3\cite{Tabarez85}, BaMnF4 \cite{Sciau90,Scott94}).  As the temperature is raised, the IC structure predominates with correlation lengths comparable to those of the long range AF ground state.  In fact, long-range incommensurate order persists in a narrow temperature range ($\approx$ 0.9 K) above the C-IC transition.    In this temperature range, third order reflections of the modulated IC structure (see Fig.\ \ref{Long_scans}(A), T = 20.99 K) and a peak close to the origin, characteristic of ferromagnetic modulations (Fig.\ \ref{Long_scans}(B)), are observed.  No evidence for modulations along any other principal direction were  observed.  

Figure\ \ref{Trans_scans} shows transverse scans at the IC peak (0,1$\pm$Q,0) with strong diffuse scattering below T$_{\rm{N}}$, signaling the onset of the incommensurate phases at elevated temperatures.  The onset for this diffuse scattering at T $\approx$ 19K correlates with the maximum observed in the the ME-effect, shown in Fig.\ \ref{ME}.  The incommensurate peaks are observed up to  T$_{{\rm CO}}\approx$ 36 K; however, in the temperature range 20.8 - to 21.70 K, they consist of two superimposed peaks, one resolution limited and the other diffuse.  This leads us to conclude that LiNiPO$_4$ undergoes two transitions, one from the short-range IC order to the long-range IC structure at T$_{IC} \cong $ 21.69, and a second at T$_{\rm{N}}$ from IC-LRO to AF as observed in the ME effect measurements.  
 
To account for the observations, two magnetic models were considered in which each spin is rotated either about the b-axis or the a-axis by an angle $\alpha$ with respect to its nearest neighbor (Model I or Model II, respectively).  The angle of the  $j$'th spin with respect to the spin at some arbitrary origin is given by $\alpha_j=\bm{Q}\cdot\bm{r}_j$, where $\bm{Q}$ is a vector along the (010) direction as the AF propagation vector. The magnetic moment in the plane varies as follows,
$\bm{S} = \mu(\sin\bm{Q}\cdot\bm{r},0,\mbox{e}^{i\bm{{\tau}}\cdot \bm{r}}\cos\bm{Q}\cdot\bm{r})$ for Model I and $\bm{S} = \mu(0, \sin\bm{Q}\cdot\bm{r}, \mbox{e}^{i\bm{{\tau}}\cdot \bm{r}}\cos\bf{Q}\cdot\bm{r})$ for Model II.  Whereas Model I predicts ferromagnetic modulations with peaks near the origin at $\bm{q} = (0,\pm Q,0)$, Model II does not.  To determine the suitable model, scans close to the origin (at small q's) along all principal directions were conducted of which only the (0,K,0) scan gave evidence to the IC structure.  Figure\ \ref{Long_scans}(B) shows background-subtracted scans along the (0,K,0) direction (background was measured at T = 40 K) with a peak observed at the intermediate long-range IC phase (between T = 20.8 K and T = 20.7 K ).  The intensities of the satellite peaks close to the AF propagation vector (see Fig.\ \ref{Long_scans}) are about a hundred times stronger than the peaks near the origin  at $(0,\pm Q,0)$, consistent with Model I as shown below.  The intensity ratio
${I(\bm{\tau}_{_{AF}} \pm \bm{Q})}/{I(\pm \bm{Q})}$ of these peaks can be estimated from the structure factors of the two reflections,
$\bm{F}_M \sim \sum_j\mbox{e}^{i\bm{q}\cdot\bm{r}_j}\hat{\bm{q}}\times(\bm{S}\times\hat{\bm{q}})$
with $\bm{q} =  \bm{\tau}_{_{AF}} \pm \mbox{n}\bm{Q};    \quad \mbox{n = 1,3}$,
yielding
\begin{equation}
\frac{I(\bm{\tau}_{_{AF}} \pm \bm{Q})} {I(\pm \bm{Q})}\approx \frac{|F(\bm{\tau}_{_{AF}} \pm \bm{Q})| ^2} {|F(\pm \bm{Q})|^2}\approx (\frac{1}{Qb})^2,
 \label{Rat_Int}
\end{equation}
where $b$ is the lattice spacing along the direction of the modulation.   Equation\ (\ref{Rat_Int}) shows the intensity of the peak near the origin vanishes as Q gets smaller, i.e., as the temperature is lowered towards the C-IC transition, giving rise to a maximum in peak intensity as qualitatively shown in Fig.\ \ref{Long_scans}.
\begin{figure}
\includegraphics[width=3.1 in]{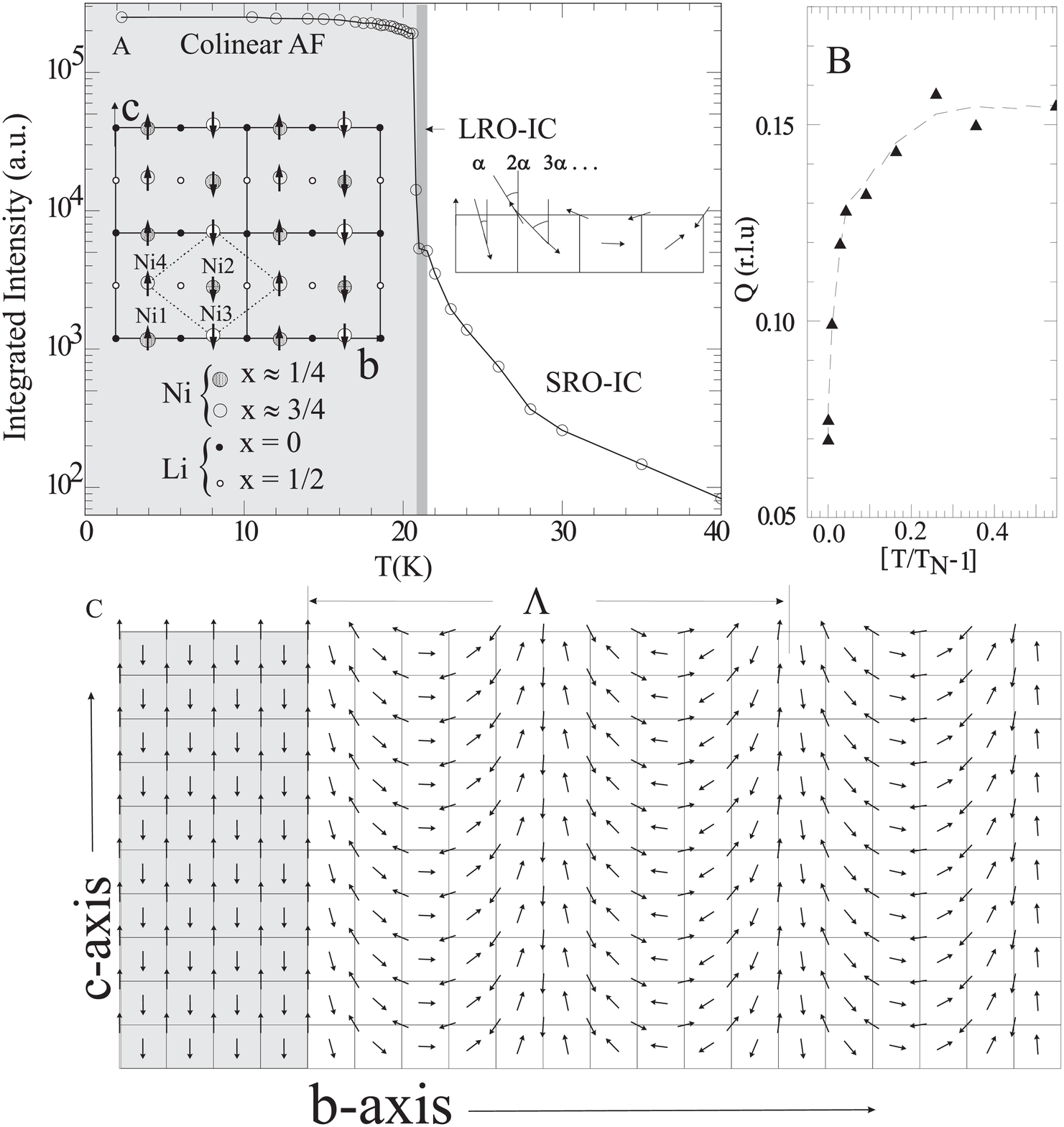}
\caption{\label{OP_Ni} A)Temperature-dependence of the order parameter in LiNiPO$_4$ as measured on the (0,1,0) magnetic Bragg reflection. Intermediate phases between the paramagnetic (at temperatures higher than $\approx$ 36 K) and the AF phases are indicated.  In the IC region two phases are identified, one with long-range and one with short-range order.  Diffuse scattering at the nominal position persists up to T$_{\rm{CO}}$ $\approx$ 36 K (a crossover temperature). The transition from short- to long-range IC order occurs at T$_{\rm{IC}}$ = 21.7K, and the C-IC transition at T$_{\rm{C-IC}} \equiv {\rm{T_N}}$ = 20.8 K.  Illustration of the ground state magnetic ordering with a projection of the relevant ions on the b-c plane is also shown.   B) Temperature dependence of the IC wave-vector. C) Simplified ground state (shaded area) beside the model of the IC structure within one plane.  Here, each spin is rotated at a finite angle with respect to a neighboring spin about the a-axis (Model II in the text).  Model I is similar to the one shown above, except the spins are rotated about the b-axis, giving rise to a FM modulated structure with a detectable peak at the origin (small angles).}
\end{figure}

The IC magnetic structure occurs as an intermediate phase between two high symmetry phases.  As the temperature is lowered from the paramagnetic phase, an onset for the IC occurs at T$_{\rm{CO}}\approx$ 36 K, with a gradual increase in the wavelength of the modulation and the correlation length.   At  T$_{\rm{IC}} $ = 21.7 K, the coherence length diverges and  higher and new harmonics appear until the modulation wavelength coincides with the high symmetry AF phase at T$_{\rm{N}}$ = 20.8 K (See Fig.\ \ref{OP_Ni}).    The origin of the IC structure may be induced by subtle charge distortions, or it could be innate to the spin Hamiltonian.    Since no evidence for a structural incommensurability was found, we hypothesize that the IC phases (long- and short-range) originate from spin-dimensionality crossover, i.e., from a continuous Heisenberg (or XY) type model to an Ising-model.  The ground state of the system is AF with no modulations, indicating a spin Hamiltonian that does not include strong terms that invoke an IC ground state (for instance, a Dzyaloshinsky-Moriya term), as recently suggested by Kharchenko {\it et al}.\cite{Kharchenko2002}.    Thus, the occurrence of intermediate magnetic incommensurate phases in LiNiPO$_4$ has all the characteristics of typical structural IC phases, which also manifest soft-phonon modes\cite{Janssen82}.  In preliminary spin wave studies of LiNiPO$_4$, a temperature-dependent gap (1.5 - 2 meV), and a minimum in the spin-wave dispersion curve was observed\cite{VakninUnpub}.  Theoretical predictions suggest that frozen magnons are possible in a 2D system with random-distribution couplings\cite{Marino2002}, as recently suggested for these systems\cite{Vaknin99,Vaknin2002}.

In summary, our observations of a C-IC first-order phase transition, and the short-range order below T$_{\rm{N}}$ explain the abrupt jump and the maximum in the ME-effect of LiNiPO$_4$ (as observed in Fig.\ \ref{ME}).  Our results however do not explain the anomalous temperature dependence of the ME coefficient which may be due to ME contributions of the spontaneous toroidal moment, which is magnetoelectric in nature\cite{Schmid2001}.   Recently, the anomalous temperature-dependence of the ME coefficient of  weakly ferromagnetic/ferroelectric/ferroelastic boracites has, for the first time, been explained by considering the contribution of the symmetry-allowed spontaneous toroidal moment\cite{Sannikov98,Sannikov2001,Sannikov2003}.  Neutron scattering studies under applied magnetic field may shed light on the anomalous behavior of the ME-effect.  We hope that our findings will stimulate theoretical microscopic studies on the nature of this IC phase observed in LiNiPO$_4$, but not in the isostructural LiCoPO$_4$.  

This work was supported (in part) under the auspices of the United States Department of Energy. The HFIR Center for Neutron Scattering is a national user facility funded by the United States Department of Energy, Office of Basic Energy Sciences- Materials Science, under Contract No. DE-AC05-00OR22725 with UT-Battelle, LLC.  The work was supported by the Department of Energy, Office of Basic Energy Sciences under contract number W-7405-Eng-82.



\end{document}